\documentclass[preprint, endfloats, showpacs, showkeywords]{revtex4}
\usepackage{epsf}
\usepackage{epsfig}
\usepackage{amsmath}
\usepackage{amssymb}
\usepackage{dcolumn}
\usepackage{graphicx}
\begin{document}
\draft


\def\be{\begin{gather}}
\def\ee{\end{gather}}
\def\VR{{\mathbf r}}                         
\def\VQ{{\mathbf q}}                         
\def\DM{\gamma\left( \VR,\VR' \right)}       
\def\half{{\scriptstyle \frac{1}{2}}}        
\def\FT{\hat{\mathcal F}}
\def\MYS{@{\hspace{0.02\linewidth}}}
\def\MYQ{q}
\def\MYVQ{{\mathbf q}}
\def\MYK{k}
\def\MYVK{{\mathbf k}}

\title{Kinetic energy density functionals for non-periodic systems}
 \author{Nicholas Choly\footnote{Electronic Address: {\tt
choly@fas.harvard.edu}} and Efthimios Kaxiras}
 \address{Department of Physics and Division of Engineering and Applied
Sciences,
	 Harvard University,\\
Cambridge, MA 02138}
\date{\today}

\begin{abstract}
 Kinetic energy functionals of the electronic density are used to model
large systems
 in the context of density functional theory, without the need to obtain
electronic
wavefunctions.
 We discuss the problems associated with the application of widely used
kinetic energy functionals
 to non-periodic systems. We develop a method that circumvents this
difficulty
and allows the kinetic energy to be evaluated entirely in real space.
 We demonstrate that the method is efficient [$O(N)$] and accurate by
comparing
the results of our real-space formulation to calculations performed in
reciprocal space, and to calculations using
traditional approaches based on electronic states.
\end{abstract}

%

%

\maketitle

Electronic structure calculations based on density functional 
theory\cite{HOHENBERG, KOHNSHAM} (DFT) and employing an
approximate kinetic energy 
 functional\cite{CARTERREVIEW, WANGTETER, PERROT, CHACON, SMARGIASSI,
WANGGOVIND, DENSDEP}
have been shown to yield accurate 
energetics for a variety of physical systems, with a considerably
smaller expenditure of computer effort than traditional schemes.
A drawback of many existing kinetic energy functionals is
 the necessity to evaluate them in part in reciprocal space. The benefits
of
 doing electronic structure calculations exclusively in real space are
several.
 Foremost among them is their ability to simulate efficiently isolated
systems,
whereas reciprocal space methods would require a large supercell 
with a significant portion of the volume devoted to uninteresting
 vacuum. A direct extension of this feature is the possibility of using
arbitrary
 boundary conditions rather than the strict periodic boundary conditions
underlying
 reciprocal space approaches; this should be of paramount importance in
dealing
 with complex structures which cannot be accommodated by simple periodic
boundary
conditions, such as dislocations, cracks, etc.
Finally, real space methods can be readily parallelized for efficient
computations on parallel computer architectures.  

In this article we examine the reasons for the 
evaluation in reciprocal space of many kinetic energy functionals and
propose a new functional form that can reproduce the energetics of
those functionals but does not require any reciprocal space evaluations.
 The performance of the new
functionals is evaluated and compared to the existing reciprocal
 space functionals as well as to the traditional approaches based on the
calculation
of electronic states.

 DFT in its usual guise consists of solving the Kohn-Sham (KS)
equations\cite{KOHNSHAM}.
The fictitious KS wavefunctions allow the exact evaluation of  
$T_s[\rho]$, the kinetic energy of
non-interacting fermions with a density $\rho(\VR)$. 
Recent calculations have
 shown that for certain systems the solution of the KS equations, usually
the
most computationally demanding part of electronic structure calculations,
can be bypassed using 
an approximate form for $T_s[\rho]$\cite{WANGGOVIND}.
 A fair number of approximate non-interacting kinetic energy functionals
have been
proposed for use within such orbital-free methods.
A subset of them are similar in 
 form\cite{WANGTETER, PERROT, CHACON, SMARGIASSI, WANGGOVIND}, sharing the
following common
traits:
\begin{enumerate}
 \item \label{item:TF} A major ingredient of $T_s[\rho]$ is the
Thomas-Fermi
(TF) energy\cite{THOMASFERMI}:
\begin{eqnarray} \label{tfenergy}
T_{TF}[\rho] = C_{TF} \int \rho^{5/3}(\VR) d\VR
\end{eqnarray}
 where $C_{TF}=\frac{3}{10} \left( 3 \pi^2 \right)^{2/3}$. Here and
throughout this
article atomic units ($\hbar = m_e = e = 1$) are employed.
 The TF energy approximates the kinetic energy in an element of space
$d\VR$ with
 that of a homogeneous non-interacting electron gas with a density
$\rho(\VR)$.
 Hence for homogeneous densities $\rho(\VR)=\rho_0$, $T_{TF}[\rho(\VR)]$
is
exact.

 \item \label{item:vW} Another important contribution is the von
Weizs\"acker
energy\cite{WEIZSACKER}, given by: 
\begin{eqnarray} \label{vwenergy}
 T_{vW}[\rho] = -\frac{1}{2} \int \sqrt{\rho(\VR)} \; \nabla^2
\sqrt{\rho(\VR)} \; d\VR
\end{eqnarray}
It can be readily shown
that the von Weizs\"acker energy yields the correct kinetic energy for a
 non-interacting fermion density that consists of a single orbital, i.e. a
one-
 or two-electron density. Also, for any density, the von Weizs\"acker
energy
 yields the energy that a system of non-interacting bosons of density
$\rho(\VR)$
would have.  This term is a lower bound of $T_s[\rho]$\cite{HERRING}.

 \item The response of a homogeneous non-interacting Fermi gas to a small
perturbation
is known exactly\cite{LINDHARD}.  This response can be related to the 
second functional derivative
 of $T_s[\rho]$ evaluated at uniform density $\rho(\VR)=\rho_0$. The
Fourier transform
 of this functional derivative is given by:
\begin{eqnarray} \label{linresponse}
\FT \left[\left. \frac{\delta^2 T_s}{\delta \rho(\VR) 
 \delta \rho(\VR')}\right|_{\rho_0} \right] = -\frac{1}{\chi_{Lind}(\MYQ)}
\\
\chi_{Lind}(\MYQ) = -\frac{k_F}{\pi^2}
 \left[ \frac{1}{2}+\frac{1-\MYQ^2}{4 \MYQ} \ln\left|
\frac{1+\MYQ}{1-\MYQ}
	\right| \right]
\end{eqnarray}
where $\chi_{Lind}(q)$ is the Lindhard response function,
 $\FT [f(\VR)]=\int f(\VR) e^{i \MYVK \cdot \VR} d\VR$ denotes the 
action of the
 Fourier transform on the function $f(\VR)$, $\MYVQ=\MYVK/2k_F$, and
$\MYQ=|\MYVQ|$.
In order to satisfy Eq.(\ref{linresponse}) in addition to 
items \ref{item:TF} and
 \ref{item:vW} above, the kinetic energy functionals under consideration
here include
 a term $T_K[\rho]$, which will be hereafter referred to as the {\em
kernel energy}:
\begin{eqnarray} \label{eqn:kernel}
 T_K[\rho] = \int f\left( \rho(\VR) \right) K(|\VR- \VR'|) g(\rho(\VR') )
d\VR \; d\VR'
\end{eqnarray}
 where for the moment $f(\rho)$, $g(\rho)$ are arbitrary functions that
can be
chosen to satisfy known limits of the exact $T_s[\rho]$.
\end{enumerate}

 The total kinetic energy functional is taken to be the sum of these
terms:
\begin{eqnarray} \label{eqn:totalfunctional}
T_s[\rho] \simeq T_{TF}[\rho]+T_{vW}[\rho]+T_K[\rho]
\end{eqnarray}
 By plugging Eq.(\ref{eqn:totalfunctional}) into Eq.(\ref{linresponse}),
it is
seen that $T_s[\rho]$ exhibits the correct linear response, provided:
\begin{eqnarray} \label{eqn:funccondition}
\lefteqn{\FT \left[\left. \frac{\delta^2 T_K}{\delta \rho(\VR) 
 \delta \rho(\VR')}\right|_{\rho_0} \right] =
-\frac{1}{\chi_{Lind}(\MYQ)}} \nonumber \\
&&   - \FT \left[\left. \frac{\delta^2 T_{TF}}{\delta \rho(\VR) 
	\delta \rho(\VR')} \right|_{\rho_0} \right] 
   - \FT \left[\left. \frac{\delta^2 T_{vW}}{\delta \rho(\VR) 
	\delta \rho(\VR')} \right|_{\rho_0} \right] 
\end{eqnarray}
 The functional derivatives of $T_{TF}$ and $T_{vW}$ are well known, and
the second
 functional derivative of $T_K$ can, by design, be easily evaluated, so
that Eq.
(\ref{eqn:funccondition}) takes the form:
\begin{eqnarray} \label{eqn:kerneleta}
\frac{2k_F}{\pi^2} f'(\rho_0) g'(\rho_0) K(\MYQ) = \hat{K}(\MYQ) \\
\label{eqn:kerneletaq}
\hat{K}(\MYQ) \equiv -\frac{k_F}{\pi^2 \chi_{Lind}(\MYQ)}-1-3 \MYQ^2 
\end{eqnarray}
 For any choice of the functions $f$ and $g$, $K(\MYQ)$ can be readily
chosen
 such that the total kinetic energy functional exhibits the correct linear
response.
 The different kinetic energy functionals considered in this paper differ
mostly in the
 choice of the functions $f$ and $g$.  

\begin{figure}
	\includegraphics[width=0.9\linewidth]{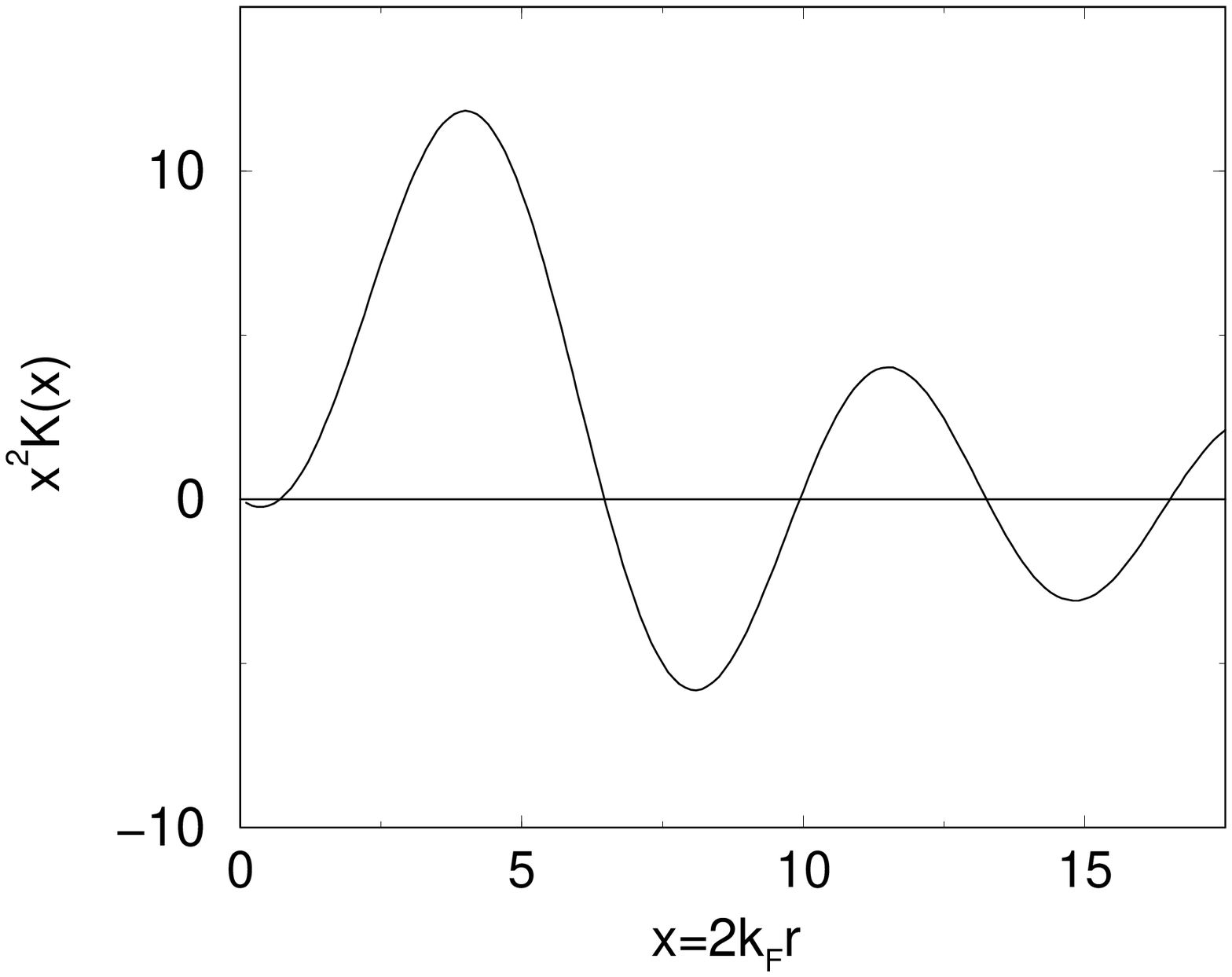}
	\caption{Kinetic energy functional kernel $K(x)$, where $x=2 k_F r$,
		multiplied by $x^2$, showing its long-ranged nature.}
	\label{fig:KR}
\end{figure}

In order to use kinetic energy functionals in actual electronic structure
calculations, an algorithm 
 must be developed for their action on discrete representations of the
charge
density.
 It is clear that $T_{TF}[\rho(\VR)]$ and $T_{vW}[\rho(\VR)]$ can be
computed easily
and efficiently on a grid in real space:
\begin{eqnarray} \label{eqn:tfcompute}
T_{TF}[\{ \rho_i \}]= \Omega C_{TF} \sum_i \rho_i^{5/3} 
\end{eqnarray}
\begin{eqnarray} \label{eqn:vWcompute}
 T_{vW}[\{ \rho_i \}] = -\frac{\Omega}{2} \sum_{ij} \sqrt{\rho_i} \;
\Delta_{ij}
	\sqrt{\rho_j}
\end{eqnarray}
 where $\Omega$ is the volume per grid point, $\{\rho_i\}$ denotes a
discrete
representation of $\rho(\VR)$ on a grid in real space, 
 and $\Delta_{ij}$ is a discrete representation of the Laplacian operator.
 In principle, we could also compute the kernel energy $T_K[\rho(\VR)]$ in
real
 space as follows:
\begin{eqnarray}
 T_K[\{ \rho_i \}]= \Omega^2 \sum_{ij} f \left( \rho_i \right) K(|\VR_j -
\VR_j|)
	g \left(\rho_j \right)
\end{eqnarray}
 However, this is impractical because of the specific form of $K(r)$.
$K(r)$ is the
 Fourier transform of $K(\MYQ)$ (see Eq.(\ref{eqn:kerneletaq})), which is
a non-trivial
Fourier transform.
Herring\cite{HERRING} has shown how it may be
 evaluated numerically. The quantity $K(r)$ is shown in Fig. \ref{fig:KR},
indicating that $K(r)$ does not decay rapidly.  In order to evaluate 
\begin{eqnarray}
\int K(|\VR-\VR'|) h(\VR') d\VR'
\end{eqnarray}
accurately, the integral needs to be evaluated over
 a sphere centered at $\VR$ of a radius equal to $\lambda$, with $\lambda$
a
large value.
 Such a convolution integral can be computed efficiently in reciprocal
space:
\begin{eqnarray} \label{eqn:qspaceconv}
 \FT \left[ \int K(|\VR-\VR'|) h(\VR') d\VR' \right] = \FT \left[ K(r)
\right]
	 \FT \left[ h(\VR) \right]
\end{eqnarray}
 Starting from $K(r)$ and $h(\VR)$ in real space, three Fast Fourier
Transforms
 (FFTs) are required to evaluate the convolution: two forward
transformations, and one
reverse transformation.
The drawback of computing the kinetic energy functional with FFTs
is that this approach maps the problem to
 a {\em periodic tiling} of the system of interest. This periodicity can
have
 consequences on the resulting physics that in many cases are undesirable
and
can lead to erroneous results.

 The problem with evaluating the kinetic energy functionals described
above in real space
lies in the
 efficient evaluation of convolution integrals. Convolution integrals with
a
long-ranged kernel can sometimes
 be evaluated efficiently in real space in an indirect way. For instance,
the
 electrostatic potential of a charge distribution $\rho(\VR)$ is the
convolution
of $\rho(\VR)$ with the very long-ranged $1/r$:
\begin{eqnarray} \label{eqn:elecpotl}
\Phi(\VR) = - \int \frac{\rho(\VR')}{|\VR-\VR'|} d\VR'
\end{eqnarray}
 The integral in this expression can be computed efficiently in real space
by
solving the Poisson equation:
\begin{eqnarray} \label{eqn:poisson}
\nabla^2 \Phi(\VR) = -4 \pi \rho(\VR),
\end{eqnarray}
In general, as an alternative to evaluating the convolution
\begin{eqnarray} \label{eqn:directconv}
V(\VR)=\int  G(\VR-\VR') f(\VR') d\VR',
\end{eqnarray}
one may equivalently solve the integral equation:
\begin{eqnarray} \label{eqn:indirectconv}
\int H(\VR-\VR')V(\VR') d\VR' = f(\VR'), \nonumber \\
H(\VR) = \FT^{-1} \left[ 
\frac{1}{\FT [G(\VR) ]} \right].
\end{eqnarray}
In discretized form, this integral equation is a standard linear problem:
\begin{eqnarray} \label{eqn:linear}
H_{ij}V_j = f_i
\end{eqnarray}
which can be solved by an iterative linear solver\cite{SAAD};
the efficiency of such approaches
 depends on the efficiency of multiplying an arbitrary vector by a matrix.
For the
 case of the Poisson equation, the matrix $H_{ij}$ becomes a discrete
representation of
 the well-localized Laplacian operator, and hence matrix-vector
multiplications
are efficient.
 However, when the integral equation that corresponds to the convolution
of
 Eq.(\ref{eqn:kernel}) is constructed, the resulting function $H_K(r)$ is
found
to be long-ranged.  The long-ranged nature of both $K(r)$ and $H_K(r)$ is
 due to the logarithmic divergence of the slope of $\chi_{Lind}(\MYQ)$ at
$\MYQ=1$,
which
causes long-ranged oscillations to appear in its Fourier transform.
Thus, matrix-vector multiplications by $H_{ij}=H_K(|\VR_j-\VR_j|)$, and 
 the solution of the integral Eq.(\ref{eqn:indirectconv}), are
inefficient.

\begin{figure}
	\includegraphics[width=0.9\linewidth]{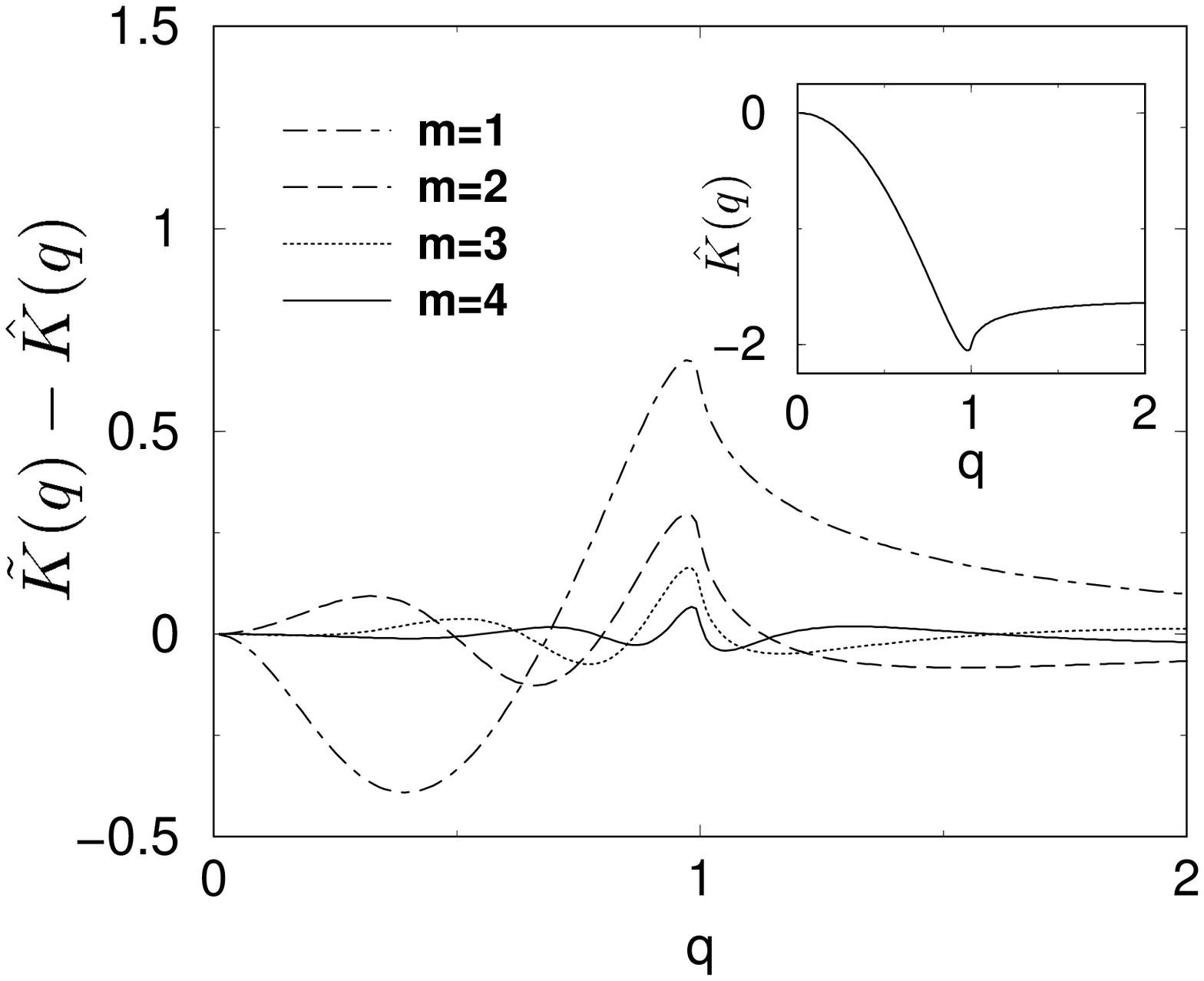}
 \caption{The difference between the rational fitting functions
$\tilde{K}(q)$
 (Eq.(\ref{eqn:rationalfn})) and $\hat{K}(q)$ (Eq.(\ref{eqn:kerneletaq}))
		with the number of terms ranging from $m=1$ to $m=4$.  The inset shows
		$\hat{K}(q)$.}
	\label{fig:rationalfit}
\end{figure}

We are therefore interested in developing a method that circumvents these
difficulties.  As a first step toward this goal, we note
 that the kernel appearing in the class of kinetic energy functionals
under consideration, $\hat{K}(\MYQ)$, can be fit well
 by the rational function:
\begin{eqnarray} \label{eqn:rationalfn}
\tilde{K}(\MYQ)=\frac{N_2 \MYQ^2 + \cdots + N_{2m} \MYQ^{2m}}
	{D_0 + D_2 \MYQ^2 + \cdots + D_{2m} \MYQ^{2m}},
\end{eqnarray}
with appropriate choices of the real coefficients $N_i$ and $D_i$.
The odd powers of $\MYQ$ are omitted because in the Taylor expansions of 
$\hat{K}(\MYQ)$ about
 $0$ and $\infty$ only even powers of $\MYQ$ appear. The quality of the
fit is shown
in Fig. \ref{fig:rationalfit} for several values of $m$.
 Next we note that $\tilde{K}(\MYQ)$ can be separated into terms of the
form:
\begin{eqnarray} \label{eqn:terms}
\tilde{K}(\MYQ)=\sum_{j=1}^m \frac{P_j \MYQ^2}{\MYQ^2 + Q_j},
\end{eqnarray}
where the $P_j$ and $Q_j$ are now complex numbers.  
With this expression, 
 the convolution of a function
$f(\VR)$ with $\tilde{K}(r)$ in reciprocal space becomes:
\begin{eqnarray} \label{eqn:ktildeconvolve}
V(\MYVQ) \equiv \tilde{K}(\MYQ)f(\MYVQ)=V_1(\MYVQ)+ \cdots + V_m(\MYVQ), 
\end{eqnarray}
where
\begin{eqnarray}
V_j(\MYVQ) = \frac{P_j \MYQ^2}{\MYQ^2 + Q_j} f(\MYVQ).
\end{eqnarray}
The $V_i(\VR)$ can be computed efficiently in real space
by solving:
\begin{eqnarray} \label{eqn:efficientvi}
(\MYQ^2 + Q_j) V_j(\MYVQ) &=& P_j \MYQ^2 f(\MYVQ) \nonumber \\
 \Rightarrow \left[ -\frac{1}{(2k_F)^2}\nabla^2 + Q_j \right] V_j(\VR) &=&
-\frac{P_j}{(2k_F)^2} \nabla^2 f(\VR)
\end{eqnarray}
that is, $V_j(\VR)$ is the solution of a complex Helmholtz equation.
 A shortcut in computing the $V_j(\VR)$ results from the fact that
$\tilde{K}(\MYQ)$,
 and thus the sum of the $V_j(\VR)$, is purely real.  For every pair of
coefficients $\{P_j,Q_j\}$, 
another pair 
 $\{P_k, Q_k\} = \{P_j^\ast, Q_j^\ast\}$ must also appear in the
expansion.
It follows that 
 $V_k(\VR)=V_j^\ast(\VR)$, and thus only half of the $V_j(\VR)$ need be
computed.
 A generalization of the kinetic energy functional with the form of
Eq.(\ref{eqn:kernel})
 has been developed by Wang et al.\cite{DENSDEP}. These functionals can
also be
treated in real space with the present method, as discussed in the
Appendix.

The following issue regarding the form of the approximation of 
Eq.(\ref{eqn:rationalfn}) deserves further
discussion:  an important feature of the Lindhard response function
 lies in the logarithmic singularity in its slope at $\MYQ=1$. As
discussed above,
this singularity manifests itself mathematically
 in the long-ranged nature of $K(r)$ and $H_K(r)$. Not surprisingly, this
singularity
 also has important physical consequences, such as Friedel oscillations
and the
 Kohn effect\cite{HARRISON}. The approximate kernel $\tilde{K}(\MYQ)$ does
not exhibit
 the singularity. It may seem then that from a physical standpoint,
$\tilde{K}(\MYQ)$
 may not adequately describe the kinetic energy of the electron gas.
However, it should
 be noted that in a discrete representation of the problem, the exact
singularity at
 $\MYQ=1$ will not be seen. Furthermore, at non-zero electronic
temperatures,
 however small, 
 the singularity in $\chi_{Lind}(\MYQ)$ disappears. Thus, one could think
of the
 fitting form $\tilde{K}(\MYQ)$ as representing $K(\MYQ)$ for a small but
finite
electronic
 temperature. The use of a fictitious, finite electronic temperature is a
trick routinely
employed to aid numerical convergence in standard DFT calculations of 
metals.

 We have therefore reduced the problem to solving the complex Helmholtz
equation,
which by itself provides a special
challenge due to the
fact that the operator is non-Hermitian.  Typical iterative methods
 for solving linear systems, like the conjugate-gradient algorithm, fail
for
non-Hermitian
 matrices. The complex Helmholtz equation is an important problem, arising
frequently
 in the context of electrodynamics. Several iterative methods have been
developed
for the special class
of complex symmetric matrices, into which complex Helmholtz 
 operators fall\cite{FREUND, COMPLEXSYMMETRIC}. For the present tests, the
biconjugate-gradient algorithm, specialized
to complex symmetric matrices\cite{FREUND}, has been employed to 
 solve Eq.(\ref{eqn:efficientvi}). Each iteration of this method requires
an amount of
computation
 that scales linearly with the system size; thus, if the number of
iterations
required to converge a solution of the Helmholtz equation does not vary 
significantly with the
 grid size, the entire method for calculating the total energy scales
linearly [$O(N)$] with
the system size (N).

\begin{figure}
	\includegraphics[width=0.9\linewidth]{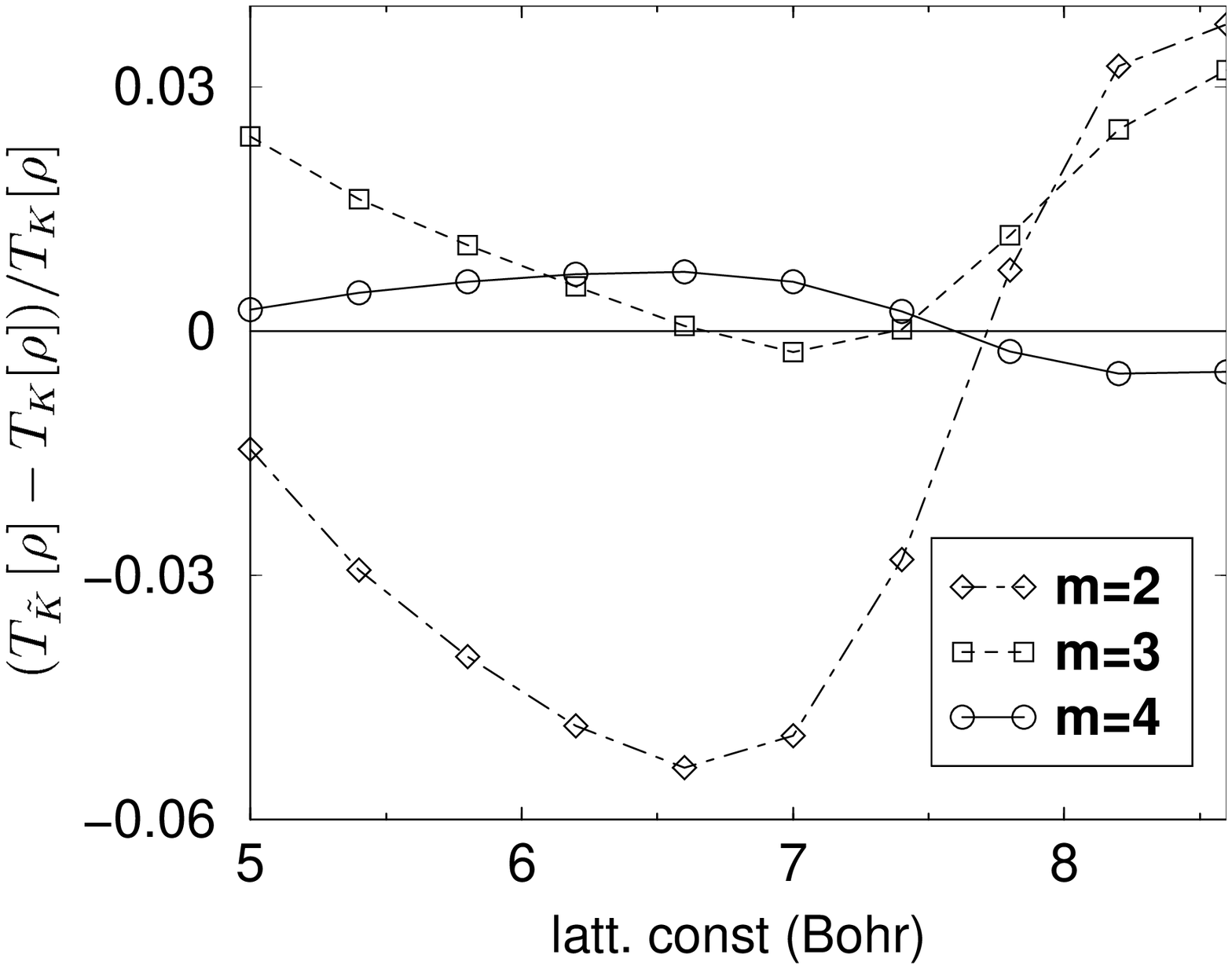}
	\caption{The fractional deviation of $T_{\tilde{K}}[\rho]$ from
		$T_K[\rho]$ for densities $\rho(\VR)$ obtained from bulk Al
		calculations (see text).}
	\label{fig:bulkal}
\end{figure}

 Solving the discretized version of Eq.(\ref{eqn:efficientvi}) in real
space suffers
from another source of inaccuracy, beyond that introduced by the fitting 
of $\hat{K}(\MYQ)$ by $\tilde{K}(\MYQ)$.  A
 discretized version of the Laplacian operator must be employed, which, in
reciprocal
 space, deviates from the exact $\MYQ^2$ behavior. The two sources of
error can be easily
 separated for the purposes of numerical tests. To measure the error due
to the
 approximation of $\hat{K}(\MYQ)$ by $\tilde{K}(\MYQ)$ alone, the kernel
energy
$T_K[\rho]$ can be computed with
the reciprocal space convolution method, Eq.(\ref{eqn:qspaceconv}), but 
using $\tilde{K}(\MYQ)$ instead
of $\hat{K}(\MYQ)$.  The error due to the 
use of the discretized Laplacian, in addition to the
fitting error, is present when the full real space evaluation method 
 is used. We found that the error due to a fourth-order discrete Laplacian
 operator was negligible compared to the error introduced by the fitting
of $\hat{K}$,
 and throughout the numerical tests this fourth-order Laplacian was
employed. The
 kinetic energy evaluated with the present real-space method is denoted by
 $T_{\tilde{K}}[\rho]$, while the kinetic energy evaluated in reciprocal
space
with $\hat{K}(q)$ is denoted by $T_K[\rho]$.
The kinetic energies $T_K[\rho]$ and $T_{\tilde{K}}[\rho]$
are evaluated and compared for a realistic set of 
 charge densities $\rho(\VR)$ in Fig. \ref{fig:bulkal}. The $T_K[\rho]$
used
is due to Wang et al.\cite{DENSDEP}, and has parameters 
$\{\alpha,\beta\}=\frac{5}{6} \pm \frac{\sqrt{5}}{6}$, and for
$T_{\tilde{K}}[\rho]$, successive
fitting orders $m=2$, 3, and 4 were tried.  The fitting coefficients for
 $m=4$ are given in Table \ref{table:PiQi}. The charge densities
considered are
\begin{table}
 \caption{Optimized fitting parameters for $\tilde{K}(\MYQ)$ with an
order-eight
 ($m=4$) rational function of Eq.(\ref{eqn:terms}). The parameters with
even
 indices $j=2, 4$ are complex conjugates of the ones given: $P_2 =
P_1^\ast$,
$P_4 = P_3^\ast$, $Q_2 = Q_1^\ast$, and $Q_4 = Q_3^\ast$.}
\label{table:PiQi}
\begin{ruledtabular}
\begin{tabular}{l|r@{\extracolsep{10pt}}r}
 & \multicolumn{1}{c}{$j=1$} & \multicolumn{1}{c}{$j=3$}  \\
\colrule
$P_j$ & $0.026696+i 0.145493$  & $-0.826696+i 0.691930$  \\
$Q_j$ & $-0.818245-i 0.370856$ & $0.343051-i 0.689646$  \\
\end{tabular}
\end{ruledtabular}
\end{table}
generated by minimizing the total energy:
\begin{eqnarray} \label{eqn:totalenergy}
 E_{tot} &=& T_{TF}[\rho] + T_{vW}[\rho] +T_K[\rho]+E_{H}[\rho] +
\nonumber \\
	&& E_{ion}[\rho]+E_{XC}[\rho]
\end{eqnarray}
 where $E_H$, $E_{XC}$, and $E_{ion}$ are the Hartree,
exchange-correlation, and
electron-ion interaction terms,
 for a bulk fcc aluminum system, with a wide range of lattice constants.
Aluminum
 was represented by the Goodwin-Needs-Heine local
pseudopotential\cite{GOODWIN},
and exchange and correlation were treated with the LDA\cite{PERDEW}.
At each lattice constant, after minimizing the electronic energy with the
kernel energy represented by $T_K[\rho]$,
the kernel energy is also computed with $T_{\tilde{K}}[\rho]$
 and compared to $T_K[\rho]$. As can be seen from Fig. \ref{fig:bulkal},
the use of
more terms in the
fitting (higher $m$) results in a smaller deviation, and at $m=4$ the
 accuracy is quite satisfactory (less than 1\% deviation for the entire
range of
lattice constants considered.)

 We have performed a different set of tests, in which at each value of the
lattice
constant the total energy with the kernel energy
represented by either  $T_K[\rho]$ or $T_{\tilde{K}}[\rho]$ is minimized 
 with respect to the density. In the test discussed earlier the density
was
fixed to that obtained from minimization of the
total energy using $T_K[\rho]$.  The present test also differs from the
 last in that the error of the discretized Laplacian is also present in
the
 evaluation of $T_{vW}[\rho]$ in the real space calculations. Thus in this
 calculation, total energy calculations are performed {\em fully and
self-consistently}
in real space, and are compared with reciprocal space results. 
 The equilibrium lattice constant and the bulk modulus obtained from these
calculations
are given in Table \ref{table:physical}.
\begin{table}
\caption{Lattice constant $a_0$, in \AA, and bulk modulus B, in GPa,
for bulk fcc aluminum with the WGC  
kinetic energy functional\cite{DENSDEP} with a density-independent
kernel, and with $\{\alpha, \beta \}=\frac{5}{6} \pm \frac{\sqrt{5}}{6}$,
compared to those values determined with the present real-space method
 with $m=2$, $3$, and $4$. Also shown is $a_0$ and B from the Kohn-Sham
(KS) calculation.}
\label{table:physical}
\begin{ruledtabular}

\begin{tabular}{r@{\extracolsep{10pt}}c@{\extracolsep{10pt}}c@{\extracolsep{10pt}}c@{\extracolsep{10pt}}c@{\extracolsep{10pt}}c}
& KS & reciprocal space & $m=4$ & $m=3$ & $m=2$ \\
\hline
$a_0$(\AA) & 4.027 & 4.035 & 4.030 & 4.045 & 4.063 \\
B (GPa) & 68.5 & 71.9 & 72.3 & 68.6 & 68.0 \\
\end{tabular}
\end{ruledtabular}
\end{table}
 As seen from this Table, $m=4$ offers an acceptable level of accuracy:
the values
 obtained with this approximation to the true kernel are essentially the
same as
 those from the exact kernel, differing only by 0.1\% for the lattice
constant
and by 0.5\% for the bulk modulus.

In conclusion, the convolution integrals that appear in the
 class of kinetic energy functionals under consideration in this paper
cannot be
 efficiently 
 evaluated directly in real space, and the corresponding inverse integral
equations
cannot be efficiently solved.  The kernels of the 
convolutions
 can be approximated by a sum of sub-kernels. This approach makes it
possible
to
evaluate the convolution efficiently in real space by solving the  
inverse integral equations that correspond to the sub-kernels, which
are complex Helmholtz equations.
 We have demonstrated that this method yields excellent results in
numerical tests,
in the sense that
 the error introduced by the real space method is negligible compared to
the error
inherent in the approximate kinetic energy functionals.

\begin{acknowledgements}
 The authors wish to thank Umesh Waghmare and Paul Maragakis for fruitful
discussions,
 and Emily Carter for a critical reading of the manuscript and useful
suggestions.
 Nicholas Choly acknowledges support from an NSF Graduate Research
Fellowship.
 This work was supported in part by a MURI-AFOSR Grant No.
F49620-99-1-0272.
\end{acknowledgements}

\appendix
\section{Real-Space Evaluation of Kinetic Energy Functionals with 
	Density-Dependent Kernels} \label{append:densdep} 

 Wang, Govind, and Carter\cite{DENSDEP} (WGC in the following) have
developed a
class of kinetic energy functionals that
 include the Thomas-Fermi and von Weizs\"acker terms, as well as a term
analogous to
 Eq.(\ref{eqn:kernel}), but with a density-dependent kernel in the
convolution:
\begin{eqnarray} \label{eqn:densdepkernel}
\lefteqn{T_K[\rho] = C_{TF} \times} \hspace{-20pt} \nonumber \\
&&\int f\left( \rho(\VR) \right) K(\rho(\VR), \rho(\VR'); 
	|\VR- \VR'|) g(\rho(\VR') ) d\VR d\VR' 
\end{eqnarray}
where
\begin{eqnarray}
 K(\rho(\VR), \rho(\VR'); |\VR - \VR'|) = K(\rho(\VR'), \rho(\VR); |\VR -
\VR'|). \nonumber
\end{eqnarray}
Numerical tests indicate that these kinetic energy functionals are more
 transferable to systems that deviate significantly from the bulk, like
surfaces.

 Even by utilizing FFTs, a straightforward evaluation of the convolution
in Eq.
(\ref{eqn:densdepkernel}) would
 require $O(N^2)$ operations, where $N$ is proportional to the size of the
system.
WGC have demonstrated how this convolution can be efficiently, 
 but approximately, evaluated. By Taylor expanding $K(\rho(\VR),
\rho(\VR'); |\VR-\VR'|)$
 with respect to $\rho(\VR)$ about some chosen average density
$\bar{\rho}$,
one obtains:
\begin{eqnarray} \label{eqn:taylorexp}
 \lefteqn{K(\rho(\VR), \rho(\VR'); |\VR-\VR'|) = K_0(|\VR-\VR'|)}
\nonumber \\
 &+& K_1(|\VR-\VR'|)\left[ \Delta \rho(\VR) + \Delta \rho(\VR')\right]
\nonumber \\
 &+& \frac{1}{2} K_{11}(|\VR-\VR'|)\left[ \Delta \rho^2(\VR) + \Delta
\rho^2(\VR')\right] \nonumber \\
&+& K_{12}(|\VR-\VR'|) \Delta \rho(\VR) \Delta \rho(\VR') + \cdots 
\end{eqnarray}
where $\Delta \rho(\VR) = \rho(\VR) - \bar{\rho}$, and
\begin{eqnarray} 
K_0(|\VR-\VR'|) &=& K(\bar{\rho}, \bar{\rho}; |\VR-\VR'|), \nonumber \\
 K_1(|\VR-\VR'|) &=& \left. \frac{\partial K(\rho(\VR), \rho(\VR');
|\VR-\VR'|)}
{\partial \rho(\VR)} \right|_{\bar{\rho}}, 	\nonumber \\
 K_{11}(|\VR-\VR'|) &=& \left. \frac{\partial^2 K(\rho(\VR), \rho(\VR');
|\VR-\VR'|)}
{\partial \rho^2(\VR)} \right|_{\bar{\rho}}, \nonumber \\
\label{eqn:taylorterms}
 K_{12}(|\VR-\VR'|) &=& \left. \frac{\partial^2 K(\rho(\VR), \rho(\VR');
|\VR-\VR'|)}
{\partial \rho(\VR) \rho(\VR')} \right|_{\bar{\rho}}, \cdots
\end{eqnarray}
 Then Eq.(\ref{eqn:densdepkernel}) can be evaluated as a sum of separate
convolutions with
kernels $K_0$, $K_1$, etc.  WGC also demonstrated that only a few terms
 of the expansion in Eq.(\ref{eqn:taylorexp}) are necessary to evaluate
the convolution
accurately for physical systems.

\begin{figure}
	\includegraphics[width=0.9\linewidth]{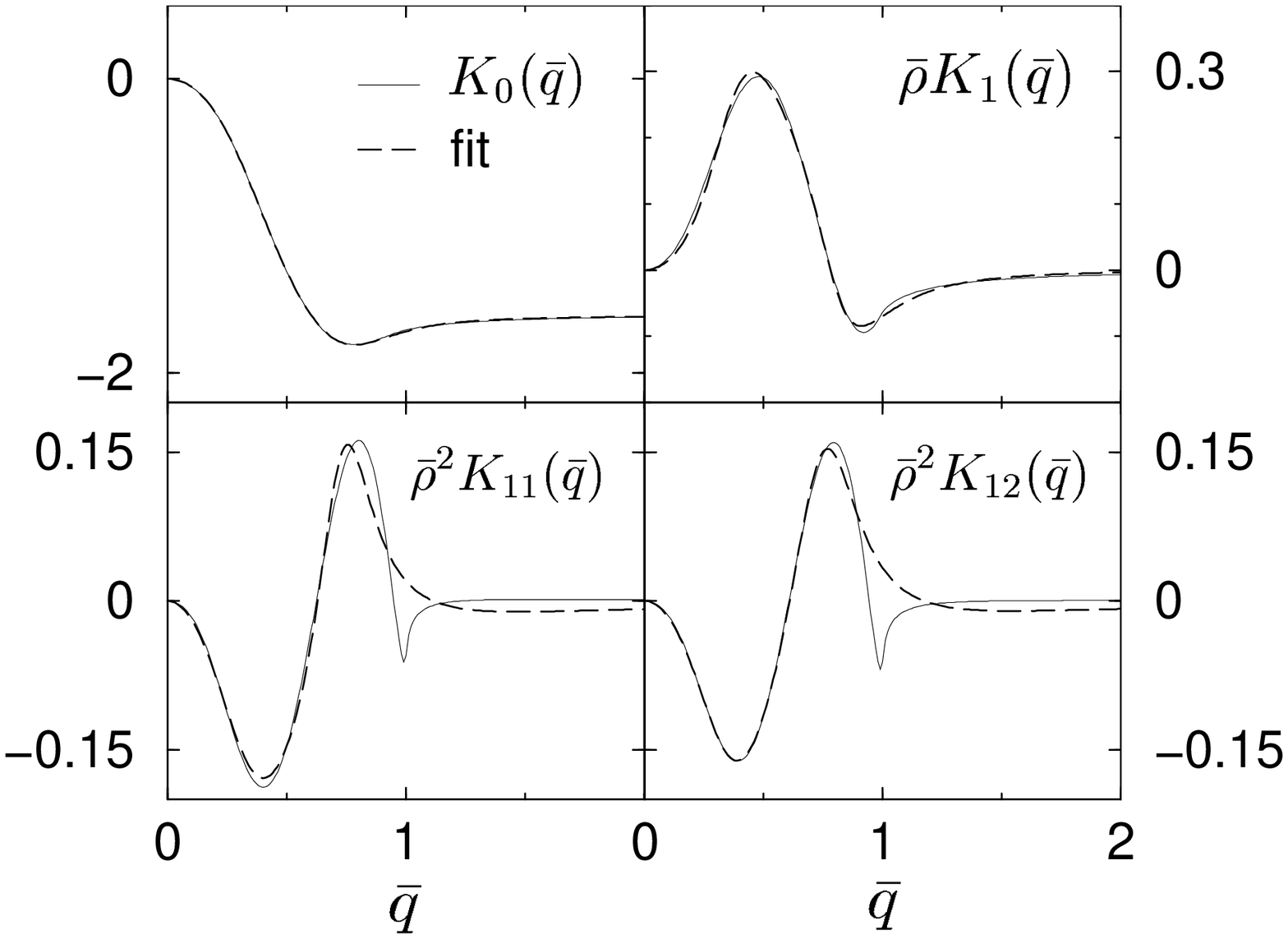}
	\caption{The fitting of the various  
	density-dependent kernels of WGC\cite{DENSDEP} by the forms of
	Eqs. (\ref{eqn:rationalfn}) and (\ref{eqn:rationalfn2}) with
 $m=4$. Notice the vastly increased scale for $K_1$, $K_{11}$, and
$K_{12}$,
 necessary to show their features, since they are negligible on the scale
	of $K_0$.}
	\label{fig:densdepfit}
\end{figure}

 The present real space method is clearly applicable to such kinetic
energy functionals,
provided $K_0(\bar{\MYQ})$, $K_1(\bar{\MYQ})$, etc., where 
$\bar{\MYQ}=k/(2\bar{k}_F)$ and $\bar{k}_F=(3 \pi^2 \bar{\rho})^{1/3}$, 
can be fit well by functions 
of the form of
Eq.(\ref{eqn:rationalfn}). For the kernels other than $K_0$, a fitting
 form slightly different than Eq.(\ref{eqn:rationalfn}) is necessary,
because
 for $\bar{\MYQ} \rightarrow \infty$, $K_0(\bar{\MYQ})$ approaches a
constant, as does the
rational form Eq.
 (\ref{eqn:rationalfn}), but the higher order kernels decay as
$\bar{\MYQ}^{-2}$. Thus
they are fit with the modified form:
\begin{eqnarray} \label{eqn:rationalfn2}
\frac{N_2 \bar{\MYQ}^2 + \cdots + N_{2(m-1)} \bar{\MYQ}^{2(m-1)}}
	{D_0 + D_2 \bar{\MYQ}^2 + \cdots + D_{2m} \bar{\MYQ}^{2m}}.
\end{eqnarray}
Functions of this form decompose into the following terms:
\begin{eqnarray} \label{eqn:terms2}
\tilde{K}(\bar{\MYQ})=\sum_{j=1}^m \frac{R_j}{\bar{\MYQ}^2 + S_j}.
\end{eqnarray}
i.e. just as in Eq. (\ref{eqn:terms}), but without the $q^2$ in the
numerator.
Now the $V_j(\VR)$ are obtained by solving a modified form of Eq.
(\ref{eqn:efficientvi}):
\begin{eqnarray} \label{eqn:efficientvi2}
 \left[ -\frac{1}{(2\bar{k}_F)^2}\nabla^2 + S_j \right] V_j(\VR) = R_j
f(\VR),
\end{eqnarray}
which is still a complex Helmholtz equation.
The kinetic energy functionals of WGC have three parameters, 
$\alpha, \beta$, and $\gamma$. Presently, only the case of 
$\{\alpha,\beta\}=\frac{5}{6} \pm \frac{\sqrt{5}}{6}$, $\gamma=2.7$,
(suggested by favorable numerical tests) is considered.
In Fig. \ref{fig:densdepfit}, the best fit to these
 kernels with a rational function of order eight ($m=4$) is shown. The
quality of
the fit is
 excellent for $K_0(\bar{\MYQ})$ and $K_1(\bar{\MYQ})$, and reasonable for
the second order kernels $K_{11}(\bar{\MYQ}), K_{12}(\bar{\MYQ})$.
WGC have shown that second order terms contribute much smaller  parts of
$T_K[\rho]$ (which is already a small part of the total energy) than the 
zeroth and first order terms, and thus a higher fraction
 of error can be tolerated in these terms. The $P_j$, $Q_j$ of the
decompositions
 of the fits to $K_0$ (Eq.(\ref{eqn:terms})) and the $R_j$, $S_j$ of the
fits
 to $K_1$, $K_{11}$, and $K_{12}$ (Eq.(\ref{eqn:rationalfn2})) are given
in Table
\ref{table:densdep}.
\begin{table}
 \caption{Optimized fitting parameters $P_j$, $Q_j$, $R_j$, and $S_j$ of
Eqs.
(\ref{eqn:terms}) and (\ref{eqn:terms2}) for fits to the kernels
 $K_0(\bar{\MYQ})$, $K_1(\bar{\MYQ})$, $K_{11}(\bar{\MYQ})$, and
$K_{12}(\bar{\MYQ})$
of the WGC density-dependent kinetic energy functional
 with $\{\alpha,\beta\}=\frac{5}{6} \pm \frac{\sqrt{5}}{6}$ and
$\gamma=2.7$.
 The parameters with even indices, $j=2,4$, are complex conjugates of the
ones
given: $X_2=X_1^\ast$, and $X_4=X_3^\ast$, where $X=P$, $Q$, $R$, or
$S$.}
\label{table:densdep}
\begin{ruledtabular}
\begin{tabular}[t]{cc|r@{\extracolsep{10pt}}r}
&  & \multicolumn{1}{c}{$j=1$} & \multicolumn{1}{c}{$j=3$}  \\
\hline
 & \raisebox{0ex}[2.5ex]{$P_j$} & $0.108403 + i 0.079657$ & $-0.908403 + i
0.439708$ \\
 \raisebox{1.2ex}[0ex][0ex]{$K_0$} & \raisebox{0ex}[0ex][1.5ex]{$Q_j$} &
$-0.470923 - i 0.465392$ & $0.066051 - i 0.259678$ \\
 & \raisebox{0ex}[2.5ex]{$R_j$} & $-0.030515 + i 0.015027$ & $0.028915 - i
0.008817$ \\
 \raisebox{1.2ex}[0ex]{$\bar{\rho} K_1$} &
\raisebox{0ex}[0ex][1.5ex]{$S_j$} & $-0.597793 - i 0.294130$ & $-0.087917
- i 0.164937$ \\
 & \raisebox{0ex}[2.5ex]{$R_j$} & $0.008907 - i 0.032841$ & $-0.034974 + i
0.009116$ \\
 \raisebox{1.2ex}[0ex]{$\bar{\rho}^2 K_{11}$} &
\raisebox{0ex}[0ex][1.5ex]{$S_j$} & $-0.537986 - i 0.233840$ & $-0.041565
- i 0.196662$ \\
 & \raisebox{0ex}[2.5ex]{$R_j$} & $0.012423 - i 0.034421$ & $-0.031907 + i
0.007392$ \\
 \raisebox{1.2ex}[0ex]{$\bar{\rho}^2 K_{12}$} &
\raisebox{0ex}[0ex][0.8ex]{$S_j$} & $-0.511699 - i 0.266195$ & $-0.034031
- i 0.188927$ \\
\end{tabular}
\end{ruledtabular}
\end{table}

\end{document}